\documentclass[twoside,twocolumn,9pt]{article}
\usepackage{extsizes}
\usepackage[super,sort&compress,comma]{natbib}
\usepackage[version=3]{mhchem}
\usepackage[left=1.5cm, right=1.5cm, top=1.785cm, bottom=2.0cm]{geometry}
\usepackage{balance}
\usepackage{widetext}
\usepackage{times,mathptmx}
\usepackage{sectsty}
\usepackage{graphicx}
\usepackage{lastpage}
\usepackage[format=plain,justification=raggedright,singlelinecheck=false,font={stretch=1.125,small,sf},labelfont=bf,labelsep=space]{caption}
\usepackage{float}
\usepackage{fancyhdr}
\usepackage{fnpos}
\usepackage[english]{babel}
\usepackage{array}
\usepackage{droidsans}
\usepackage{charter}
\usepackage[T1]{fontenc}
\usepackage[usenames,dvipsnames]{xcolor}
\usepackage{setspace}
\usepackage[compact]{titlesec}
\usepackage{epstopdf}%This line makes .eps figures into .pdf - please comment out if not required.

\definecolor{cream}{RGB}{222,217,201}

\begin{document}

\pagestyle{fancy}
\thispagestyle{plain}
\fancypagestyle{plain}{

%%%HEADER%%%
%\fancyhead[C]{\includegraphics[width=18.5cm]{head_foot/header_bar}}
%\fancyhead[L]{\hspace{0cm}\vspace{1.5cm}\includegraphics[height=30pt]{head_foot/journal_name}}
%\fancyhead[R]{\hspace{0cm}\vspace{1.7cm}\includegraphics[height=55pt]{head_foot/RSC_LOGO_CMYK}}
%\renewcommand{\headrulewidth}{0pt}
}
%%%END OF HEADER%%%

%%%PAGE SETUP - Please do not change any commands within this section%%%
\makeFNbottom
\makeatletter
\renewcommand\LARGE{\@setfontsize\LARGE{15pt}{17}}
\renewcommand\Large{\@setfontsize\Large{12pt}{14}}
\renewcommand\large{\@setfontsize\large{10pt}{12}}
\renewcommand\footnotesize{\@setfontsize\footnotesize{7pt}{10}}
\makeatother

\renewcommand{\thefootnote}{\fnsymbol{footnote}}
\renewcommand\footnoterule{\vspace*{1pt}%
\color{cream}\hrule width 3.5in height 0.4pt \color{black}\vspace*{5pt}}
\setcounter{secnumdepth}{5}

\makeatletter
\renewcommand\@biblabel[1]{#1}
\renewcommand\@makefntext[1]%
{\noindent\makebox[0pt][r]{\@thefnmark\,}#1}
\makeatother
\renewcommand{\figurename}{\small{Fig.}~}
\sectionfont{\sffamily\Large}
\subsectionfont{\normalsize}
\subsubsectionfont{\bf}
\setstretch{1.125} %In particular, please do not alter this line.
\setlength{\skip\footins}{0.8cm}
\setlength{\footnotesep}{0.25cm}
\setlength{\jot}{10pt}
\titlespacing*{\section}{0pt}{4pt}{4pt}
\titlespacing*{\subsection}{0pt}{15pt}{1pt}
%%%END OF PAGE SETUP%%%

%%%FOOTER%%%
\fancyfoot{}
%\fancyfoot[LO,RE]{\vspace{-7.1pt}\includegraphics[height=9pt]{head_foot/LF}}
%\fancyfoot[CO]{\vspace{-7.1pt}\hspace{13.2cm}\includegraphics{head_foot/RF}}
%\fancyfoot[CE]{\vspace{-7.2pt}\hspace{-14.2cm}\includegraphics{head_foot/RF}}
%\fancyfoot[RO]{\footnotesize{\sffamily{1--\pageref{LastPage} ~\textbar  \hspace{2pt}\thepage}}}
%\fancyfoot[LE]{\footnotesize{\sffamily{\thepage~\textbar\hspace{3.45cm} 1--\pageref{LastPage}}}}
%\fancyhead{}
\renewcommand{\headrulewidth}{0pt}
\renewcommand{\footrulewidth}{0pt}
\setlength{\arrayrulewidth}{1pt}
\setlength{\columnsep}{6.5mm}
\setlength\bibsep{1pt}
%%%END OF FOOTER%%%

%%%FIGURE SETUP - please do not change any commands within this section%%%
\makeatletter
\newlength{\figrulesep}
\setlength{\figrulesep}{0.5\textfloatsep}

\newcommand{\topfigrule}{\vspace*{-1pt}%
\noindent{\color{cream}\rule[-\figrulesep]{\columnwidth}{1.5pt}} }

\newcommand{\botfigrule}{\vspace*{-2pt}%
\noindent{\color{cream}\rule[\figrulesep]{\columnwidth}{1.5pt}} }

\newcommand{\dblfigrule}{\vspace*{-1pt}%
\noindent{\color{cream}\rule[-\figrulesep]{\textwidth}{1.5pt}} }

\makeatother
%%%END OF FIGURE SETUP%%%

%%%TITLE, AUTHORS AND ABSTRACT%%%
\twocolumn[
  \begin{@twocolumnfalse}
\vspace{3cm}
\sffamily
\begin{tabular}{m{4.5cm} p{13.5cm} }

 & \noindent\LARGE{\textbf{Current-Voltage Characteristics of Borophene and Borophane Sheets}} \\%Article title goes here instead of the text "This is the title"
\vspace{0.3cm} & \vspace{0.3cm} \\

 & \noindent\large{Sahar Izadi Vishkayi$^{\textit{a}}$, and Meysam Bagheri Tagani$^{\ast}$\textit{$^{a}$}} \\%Author names go here instead of "Full name", etc.

 & \noindent\normalsize{Motivated by recent experimental and theoretical research on a monolayer of Boron atoms, Borophene, current-voltage characteristics of three different Borophene sheets, 2Pmmn, 8Pmmn and 8Pmmm, are calculated using density functional theory combined with nonequilibrium Green function formalism. Borophene sheets with two and eight atoms in a unit cell are considered and bandstructure, electron density and structural anisotropy of them are analyzed in details. Results show that 8Pmmn and 8Pmmm structures which have eight atoms in the unit cell have less anisotropy than 2Pmmn. In addition, although 8Pmmn shows a Dirac cone in the bandstructure, its current is lower
 than two others. We also consider a fully hydrogenated Borophene, Borophane, and find that the hydrogenation process reduces the structural anisotropy and the current significantly. Our findings reveal that the current-voltage characteristics of the Borophene sheets can be used to detect the kind and the growth direction of the sample because it is strongly dependent on the direction of the electron transport, anisotropy and details of the unit cell of the Borophene.} \\

\end{tabular}

 \end{@twocolumnfalse} \vspace{0.6cm}

  ]
%%%END OF TITLE, AUTHORS AND ABSTRACT%%%

%%%FONT SETUP - please do not change any commands within this section
\renewcommand*\rmdefault{bch}\normalfont\upshape
\rmfamily
\section*{}
\vspace{-1cm}

%%%FOOTNOTES%%%

\footnotetext{\textit{$^{a}$}~Address, Department of Physics, Computational Nanophysics Laboratory (CNL), University of Guilan, Po Box:41335-1914, Rasht, Iran; E-mail: $m{\_}bagheri@guilan.ac.ir$}

\section{Introuduction}
\par Discovery of graphene \cite{graphene} opens a window for fabrication of two-dimensional (2D) materials which possibility of their synthesis was not predicted before. Graphene has attracted a lot of attention in recent years due to its unique properties such as high electron and thermal conductivity, high mechanical strength, and chemical inertness \cite{grproperties,grElectric,grTransparentElectrode,grThermal,grMechanicallyStrong,grIntrinsic}. After successful synthesis of Graphene, scientists searched both experimentally and theoretically to obtain new 2D materials. Silicene- a silicon analogue to Graphene- was successfully synthesized by Vogt et al. in 2012 \cite{silicene}. In 2014, the first report about the synthesis of Germanene- a Germanium analogue to Graphene- was published by Davila and coworkers\cite{germanene}. One year later, Zhu et al. synthesized a monolayer of Tin, Stanene, under ultra-high vacuum condition at room temperature \cite{stanine}.

\par Mannix and co-workers synthesized Borophene, a 2D-Boron sheet, on Ag (111) under ultrahigh vacuum condition in 2015 \cite{Mannix}. The synthesized structure exhibited a strong anisotropy, furthermore, mechanical properties of the sample were substantial. At the same time, Feng et.al \cite{feng} synthesized 2D-Boron sheets by using molecular beam epitaxy method and observed $\chi_3$ and $\beta_{12}$ phases \cite{Wu} with triangular lattices. Nonetheless, there is a fundamental difference between these outstanding works: the obtained structure by Mannix is corrugated, whereas, the structure prepared in the second work was completely planner. According to the results of Ref. \cite{Mannix}, it was predicted that the freestanding Borophene has space group of Pmmn and two atoms in the unit cell introduced as 2Pmmn. Freestanding Borophene is not stable due to the existence of a soft mode in long wavelength in its phonon band structure \cite{Mannix, Peng1, Peng2}. Zhou and co-workers \cite{Zho} showed that two freestanding allotropes of the Borophene, as 8Pmmn and 8Pmmm, can be stable. 8Pmmn Borophene has massless Dirac fermions which make it similar to graphene but with anisotropy Fermi velocity of electrons. Afterward, it was revealed that the 8Pmmn Borophene can be considered as the first single element two-dimensional structure with ionic bonding \cite{Benzaliha}. Borophene has attracted a lot of attention in last year and its optical properties \cite{Peng2, ref1}, electronic properties of its nanoribbons \cite{ref2}, transport properties \cite {ref3}, oxidization effect~\cite{ref1, ref4}, surface hydrogenation ~\cite{ref5, ref6}, superconductivity \cite{ref7, ref8, ref9}, mechanical properties \cite {ref10, ref11}, phonon transport \cite {ref12, ref13}, and its application in batteries \cite {ref14, ref15, ref16, ref17} have been searched recently.

\par In this article, we study the electronic properties of four 2D-Boron structures as corrugated Borophene, 2Pmmn, fully hydrogenated Borophene, 2HPmmn, stable Borophene with Dirac cones, 8Pmmn, and a bilayer Borophene sheet known as 8Pmmm. Their structure differences are examined using density functional theory and their responses to external bias voltages are analyzed using nonequilibrium Green function formalism. To explore the role of the structural anisotropy in details, we consider two different scenarios for the current. Indeed, the currents in x- and y-directions are computed and their differences are investigated.  Results show that how anisotropy in the structure affects the current-voltage characteristics of the samples. We show that the current of the sample can be used as an analysis tool to verify the growth direction of the sample and even the kind of the sample. Section 2 is devoted to the method of the simulation and parameters used to optimize the structures and calculate the current. Results are presented in section 3. Some sentences are given as a summary in the end.

%\subsubsection{This is the subsubsection style.~~} These headings should end in a full point.

%\paragraph{This is the next level heading.~~} For this level please use \texttt{\textbackslash paragraph}. These headings should also end in a full point.
\section{Method}
\par SIESTA package \cite{SIESTA} was used to optimize and compute electronic properties of the considered structures. Perdew-Burke-Ernzerhof (PBE) \cite{PBE} generalized gradient approximation (GGA) was employed as exchange-correlation functional and norm-conserved Troullier-Martins pseudopotentials \cite{Troulliner} were applied to describe the interaction between the valence and core electrons. A double zeta-single polarized basis set (DZP) was used and density mesh cut-off was set to $200 Ry$. The integration in the k-space was performed using a $60 \times 60 \times 1$ Monkhorst- Pack k-point mesh centered at $\Gamma$-point. The forces on all atoms were less than  $10^{-3} eV/${\AA}, and $20$ {\AA} vacuum was considered  to eliminate interlayer interactions.
\par Transport properties of the structures were studied using nonequilibrium Green function formalism. The current was computed using Landauer-B\"{u}ttiker formula as
\begin{equation}
I=\frac{2q}{\hbar} \int dE T(E) (f_{L}(E)-f_{R} (E)),
\end{equation}
where $f_{\alpha}(E)=\frac{1}{1+exp(\frac{E-\mu_\alpha}{k_BT})}$ is Fermi-Dirac distribution function of $\alpha$-electrode,  $\mu_{L(R)}=E_f\pm qV/2$ describes the chemical potential of left (right) lead and $V$ denotes the external bias voltage. $T(E)=Tr(\Gamma_L(E)G^A(E)\Gamma_R(E)G^R(E))$ is the transmission coefficient of the structure \cite{Datta} which is dependent on the coupling geometry, shape of the central region, and external bias. To compute the transmission coefficient, TRANSIESTA code \cite{TRANSIESTA} was used. We set $1400$ k-points perpendicular to the transmission direction and $120$ k-points parallel to the transport direction to take into account the two-dimensional structures of Borophene and Borophane perfectly.

\section{Results and Discussions}
Fig.1 shows four considered structures. 2Pmmn has two atoms in a unit cell that are placed in two different plans, shown by red and blue balls in Fig.1 (a). According to the Mulliken analysis, no electron transfers between the Boron atoms in the unit cells, so they are coupled to each other by covalent bonding.
\begin{figure}[h]
\centering
  \includegraphics[height=80mm,width=80mm,angle=0]{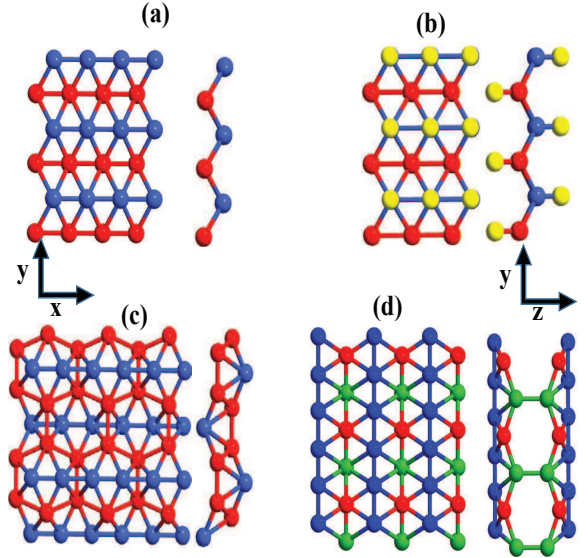}%%
  \caption{The schematic of in-plane (left) and side (right) views of four considered structures. a) 2Pmmn, b)2HPmmn, c)8Pmmn, and d)8Pmmm. The left (right) axis shows the direction of atoms in in-plan (side) view. (color on-line) }
  \label{Fig1}
\end{figure}

Fig.1 (b) shows 2HPmmn structure which Hydrogen atoms are in opposite side of the adjacent horizontal line of Boron atoms. The hydrogenation of Borophene increases the lattice vector along a direction, see table 1, about $17$ percent
which is consistent with results of ref. \cite{ref6} and reduces the buckling height. From the Mulliken analysis, we find that $0.1$ electron charge is transferred in B-H bonding. In Fig.1 (c), another structure with high symmetrical space group Pmmn and 8 Boron atoms in a unit cell is drown. 8Pmmn Borophene was predicted to be stable in freestanding condition \cite{Zho}. Bezanilla et al. \cite{Benzaliha} showed that the Boron atoms are classified into two types, inner atoms (red balls) and ridge atoms (blue balls). The ridge atoms gain negative charge, about 0.14 e, thus, the inner and ridge atoms are placed next to each other by ionic bonding.  Another structure with 8 Boron atoms in a unit cell with high symmetry space group Pmmm is shown in Fig.1 (d) and known as 8Pmmm. Three types of Boron atoms exist in the structure, ridge atoms ($B_R$, blue balls), middle atoms ($B_{m}$, red balls) and inner atoms ($B_{I}$, green balls). $B_I$ atoms are pillars of the structure jointing two planes of the Borophene, so that the structure can be thought of as a hexagonal bilayer Borophene. A similar structure has been recently introduced as P6/mmm \cite{ref18}. Mulliken analysis shows that the $B_R-B_{m}$ bonding is covalent whereas, $B_R-B_{I}$ bonding is ionic.  0.15e is gained by each $B_R$ atoms while each $B_{I}$ loses 0.3e. Lattice constants (a,b), buckling height (h), total energy ($E_{cell}$) and binding energy ($E_B=\frac{E_{free  atom}-E_{cell}}{N}$, where $N$ is the number of atoms in the unit cell) of the considered structures are listed in table 1. The data shows that the 8Pmmm structure has the lowest binding energy which is in good agreement by the achievements of ref.~\cite{Zho}. 2Pmmn has lower energy than 2HPmmn. The Borophene synthesized by Mannix et.al. \cite{Mannix} was just partially hydrogenated over times which can be explained by the different binding energy between 2Pmmn and 2HPmmn.

\begin{table}[h]
\small
  \caption{\ The structural properties of the considered structures. a and b are the lattice vectors in x and y direction, respectively, h is the buckling length}
  \label{tbl1}
  \begin{tabular*}{0.5\textwidth}{@{\extracolsep{\fill}}cccccc}
    \hline
    Structure & a({\AA}) & b({\AA}) & h({\AA}) & total energy(eV) & binding energy($\frac{eV}{atom}$)  \\
    \hline
    2Pmmn & 1.6 & 2.82 & 0.91 & -158.25914 & -6.51559 \\
    2HPmmn & 1.87 &	2.84 &	0.84 &	-190.27067 &-5.0759834 \\
    8Pmmn & 3.26 & 4.52	& 2.25 & -634.66187 & -6.718754 \\
    8Pmmm & 2.89 & 3.28	& 4.15 & -634.94009 & -6.753531 \\
    \hline
  \end{tabular*}
\end{table}

\par Bandstructures and density of states (DOS) of the structures are drawn in Fig. 2. 2Pmmn is an anisotropic metal because three bands cross the Fermi level in $a$ direction (2 bands along G-X and 1 band along Y-S) while it has a big gap in the corrugated direction (along G-Y and X-S). Our results are consistent with previously published works~\cite {Mannix, Peng1,ref1} except ~\cite{ref3} that found four metallic bands. The metallic behavior of 2Pmmn is completely supported by the DOS graph.  At $E-E_F=-1 eV$, a van Hove singularity is observed in the DOS attributed to the saddle point of the band structure in the $\Gamma$ point~\cite{ref1}. Dirac cone is observed in the band structure of 2HPmmn. Although, the electronic configuration of $B-H$ and its band structure look like Graphene, there are some significant differences between them \cite{ref6}. The Dirac cone of 2HPmmn is anisotropic because of the difference of the band slopes in two sides of the Dirac point. Moreover, the velocity of Fermi electrons in 2HPmmn is 2-4 times more than Graphene ones. Xu et al.~\cite{ref6} showed that the Dirac cone of 2HPmmn comes from the hybridization of $p_x$ and $p_y$ orbitals, but $p_z$ orbitals are responsible for the Dirac cone of Graphene. The presence of the Dirac point is sustained by the DOS at $E-E_F=0$.
\begin{figure}[h]
\centering
  \includegraphics[height=80mm,width=100mm,angle=0]{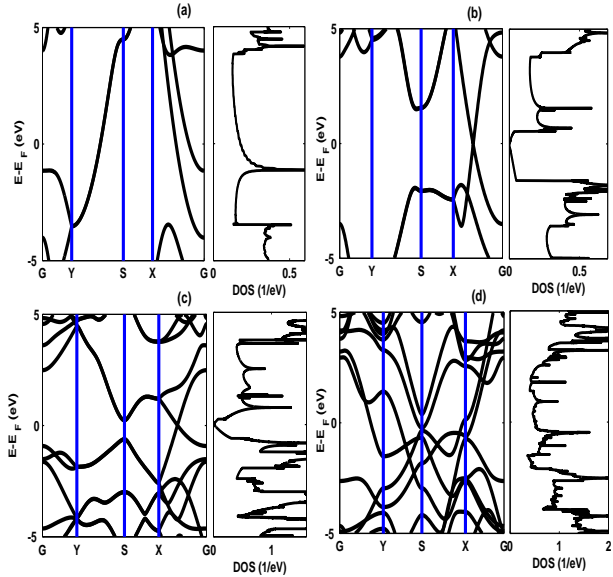}%%
  \caption{The bandstructures and density of states of a) 2Pmmn, b) 2HPmmn, c)8Pmmn, and d) 8Pmmm. ) }
  \label{Fig2}
\end{figure}

\par 8Pmmn is a gapless semi-metal, see Fig.2 (c) which has an anisotropic Dirac cone along G-X direction. There are some differences between the Dirac cone of 8Pmmn and 2HPmmn as follows: 1- The Dirac cone slope of 8Pmmn is lesser than 2HPmmn. 2- Fermi electrons of 8Pmmn are slower than 2Pmmn ones. 3-More electronic states exist near the Fermi level in 8Pmmn (see the DOS or the band structure in S point). 4- The origin of the Dirac cone of 8Pmmn is attributed to the $p_z$ orbitals of inner Boron atoms that make disorder hexagonal pattern \cite{Benzaliha}. Fig.2 (d) shows the band structure and DOS of 8Pmmm. It is obvious that 8Pmmm is an isotropic metal with a lot of electronic states around the Fermi level.  Metallic behavior of the structure is supported by the DOS. No gap is observed in its DOS spectrum.
\begin{figure}[h]
\centering
  \includegraphics[height=50mm,width=80mm,angle=0]{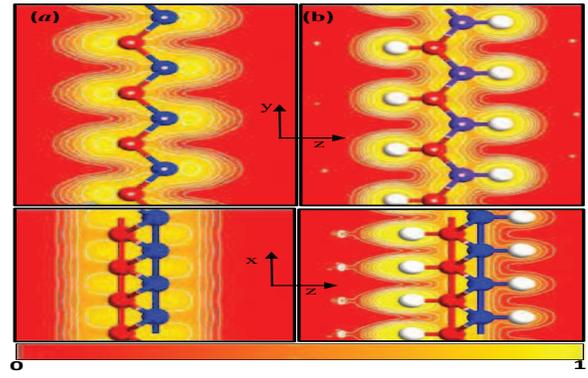}%%
  \caption{The electron localization function of a) 2Pmmn, and b) 2HPmmn.}
  \label{Fig3}
\end{figure}

\par Electron localization function (ELF) of 2Pmmn and 2HPmmn is drawn in Fig.3. It is clear that in corrugation direction (along y), the electrons are completely localized, while, localization is weaker for 2Pmmn structure. This strong localization of the electrons will lead to suppression of the current for 2HPmmn in y direction (discussed later). In x direction, the electron is delocalized in both structures and $ELF=0.5$ can be interpreted as the quasi-free electron. Thus, we expect that two structures behave as a metal in x direction.

\begin{figure}[h]
\centering
  \includegraphics[height=80mm,width=100mm,angle=0]{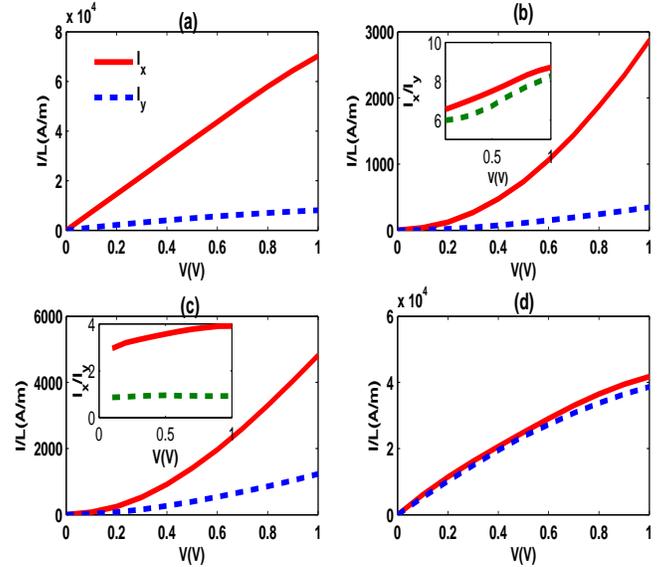}%%
  \caption{The current-voltage characteristics of a) 2Pmmn, b) 2HPmmn (the inset shows the $\frac{I_x}{I_y}$ for 2Pmmn (solid line) and 2HPmmn (dashed line)), c)8Pmmn (the inset shows the $\frac{I_x}{I_y}$ for 8Pmmn (solid line) and 8Pmmm (dashed line)) , and d) 8Pmmm.  }
  \label{Fig4}
\end{figure}

\par In order to more analyze the role of structural anisotropy of the samples, the current-voltage characteristic of the four considered structures is shown in Fig.4. For each structure, we consider two scenarios: the external bias voltage is applied along x-direction or along y-direction. In all cases, the current per length of the central region is computed. The current-voltage characteristic of  2Pmmn is ohmic, Fig.4a, but its slope is direction-dependent so that the slope is more for $I_x$. The bonding length along x-direction is shorter than y-one, therefore, the electron transport is more desirable in this direction. In the other word, the resistivity of the 2Pmmn structure is anisotropic like its mechanical properties \cite{ref11}.

\par The current-voltage characteristic of 2HPmmn shows anisotropy. The current is more in x-direction which is in good agreement with ELF prediction. In addition, it is clear that the magnitude of the current reduces significantly when the Borophene is hydrogenated. The transition from a metal, 2Pmmn, to a semi-metal, 2HPmmn, significantly reduces the current. Unlike 2Pmmn case, the current-voltage characteristic of 2HPmmn is not linear which is directly related to its bandstructure. Although $\frac{I_x}{I_y}$ increases with voltage, the ratio is moderated by hydrogenation, see the inset of Fig.4 b. It comes from the increase of the lattice constant in x-direction. Ref.~\cite{ref3} studied the same structure and found that hydrogenation reduces the anisotropy in very low bias voltages. Here, we considered high voltage regime and found that the structural anisotropy increases by voltage.
\begin{figure}[h]
\centering
  \includegraphics[height=80mm,width=100mm,angle=0]{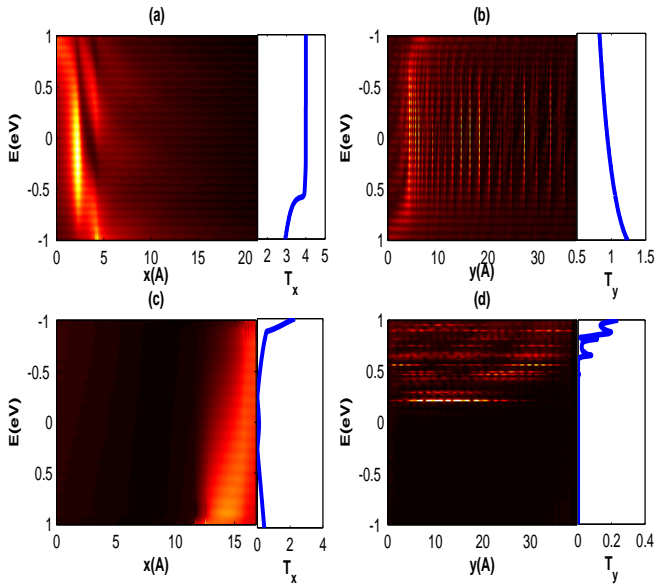}%%
  \caption{The projected local density of states and transmission coefficients of a) 2Pmmn along x-direction, b) 2Pmmn along y-direction, c)2HPmmn along x-direction , and d) 2HPmmn along y-direction at $V=0.5V$. }
  \label{Fig4}
\end{figure}
\par Local density of states (LDOS) projected along transport direction and transmission coefficient at $V=0.5V$ are shown in Fig.5 for 2Pmmn and 2HPmmn. The transmission coefficient in the x-direction is four times more than in the y-one leading to a significant increase of the current along x in 2Pmmn case. This phenomenon comes from the strong coupling of the Boron atoms along x-direction as it was shown in the ELF analysis. It is obvious that the LDOS is more in the left side of the central channel which is coupled to the source. In addition, it is observed that the coupling of the transport channel to the source is stronger in the x-direction. One can also observe that the LDOS of y case is nearly discontinuous along the transport direction attributed to the long bonding length in the y-direction. About 2HPmmn, although the appearance of the transmission coefficient is the same in both directions, there are some differences between them. Results show that the transmission coefficient is zero in $\epsilon=\mu_L$ and $\epsilon=\mu_R$ and increases between them by the increase of the voltage resulting in the nonlinear increase of the current. LDOS analysis shows that the electron states are more near the drain, right side. In addition, the LDOS spectrum exhibits there are more electron states when the transport is in the x-direction.

\par The current-voltage characteristic of 8Pmmn is also dependent on the direction of the transport channel. Unlike 2Pmmn case, the ratio of $I_x$  to $I_y$ is decreased by increase of the voltage so that $I_x/I_y$ becomes constant in high voltages, inset of Fig.4 c. The current of the structure is not linear because it is a semimetal. It is found that the ridge atoms have a better participation in the transport than the inner ones, so $I_x>I_y$. Although both 8Pmmm and 2HPmmn have a Dirac cone, the current of 8Pmmn is more than 2HPmmn. This phenomenon is related to the small band gap of 8Pmmn in $S$ point of the brillouin zone. Unlike three previous cases, the current of 8Pmmm is nearly isotropic so that $\frac{I_x}{I_y}\simeq 1$. Indeed, the results show that the transition from a Borophene with two atoms in the unit cell to one with eight atoms significantly shrinks the anisotropy of the structure.  There are significant differences between the current of four considered structures, hence, the current-voltage analysis can be used as a good tool to verify the structure of the synthesized Borophene.

\begin{figure}[h]
\centering
  \includegraphics[height=50mm,width=100mm,angle=0]{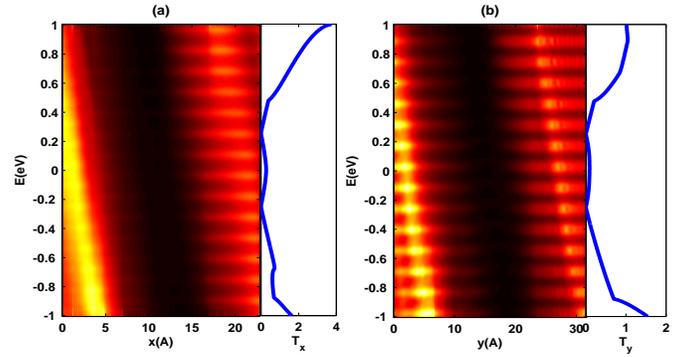}%%
  \caption{The projected density of states and the transmission coefficients of 8Pmmn a) along x-direction and b) along y-direction at $V=0.5V$. }
  \label{Fig4}
\end{figure}

\par The projected local density of states (LDOS) of 8Pmmn structure is plotted in Fig.6. It is observed that there are more electron states in the bias window when the transport channel is in the x-direction. Furthermore, $T_x$ is more than $T_y$ so $I_x>I_y$. To specify the contribution of the inner, ridge and middle atoms in the current of the 8Pmmm structure, density of states (DOS) is plotted in Fig.7 for $V=0.5V$. It is found that the ridge atoms participate in the transport more than the others. In addition, it is clear that the inner and middle atoms have equal weight in the charge transport.

\begin{figure}[h]
\centering
  \includegraphics[height=50mm,width=90mm,angle=0]{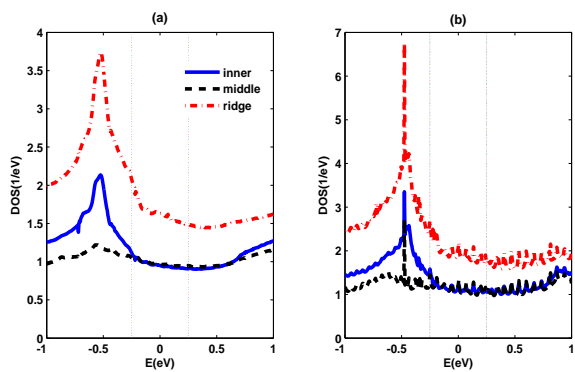}%%
  \caption{The density of states of inner (solid lines), middle (dashed lines), and ridge (dash-dotted lines) atoms in 8Pmmm along a) x-direction and b) y-direction at $V=0.5V$. Dotted lines show bias window. }
  \label{Fig4}
\end{figure}

\section{Conclusions}
We have studied the current-voltage characteristics of three different Borophene sheets using density functional theory combined with nonequilibrium Green function formalism. Results show that the increase of the Boron atoms in the unit cell of the Borophene sheet significantly reduces the structural anisotropy. In addition, a transition from  covalent to ionic bonding is observed in the Borophene by increase of the unit cell atoms. Structural anisotropy directly affects on the current of the sample so that the current is dependent on the direction of the transport. A fully hydrogenated Borophene as Borophane is also considered. Our findings show that the hydrogenation can significantly moderate the anisotropy but it also reduces the current of the sample. Furthermore, it is observed that the Borophene sheets with Dirac cone have lower current than the others.

\end{document}